\documentclass[aps,prb,reprint,superscriptaddress]{revtex4-2}
\usepackage{hyperref}
\usepackage{graphicx}
\usepackage{color}
\usepackage{amsfonts}
\usepackage{amsmath}
\usepackage[utf8]{inputenc}
\usepackage[T1]{fontenc}
\usepackage{mathtools}
\usepackage{bbm}
\usepackage[dvipsnames]{xcolor}
\hypersetup{colorlinks=true, urlcolor=blue, citecolor = blue}

\begin{document}

\title{Scanning gate microscopy probing of anisotropic electron flow\\ in a two dimensional electron gas at the (110) $\mathrm{LaAlO}_3/\mathrm{SrTiO}_3$ interface:\\ A theoretical investigation}

\author{M. P. Nowak}        
\email{mpnowak@agh.edu.pl}
\affiliation{AGH University of Krakow, Academic Centre for Materials and Nanotechnology, al. A. Mickiewicza 30, 30-059 Krakow, Poland}

\author{M. Zegrodnik}        
\email{michal.zegrodnik@agh.edu.pl}
\affiliation{AGH University of Krakow, Academic Centre for Materials and Nanotechnology, al. A. Mickiewicza 30, 30-059 Krakow, Poland}

\author{D. Grzelec}
\affiliation{AGH University of Krakow, Faculty of Physics and Applied Computer Science, al. A. Mickiewicza 30, 30-059 Krakow, Poland}

\author{B. Szafran}
\affiliation{AGH University of Krakow, Faculty of Physics and Applied Computer Science, al. A. Mickiewicza 30, 30-059 Krakow, Poland}

\author{R. Citro}
\affiliation{Department of Physics E. R. Caianiello University of Salerno and CNR-SPIN, Via Giovanni Paolo II, 132, Fisciano (Sa), Italy}

\author{P. Wójcik}
\affiliation{AGH University of Krakow, Faculty of Physics and Applied Computer Science, al. A. Mickiewicza 30, 30-059 Krakow, Poland}

\date{\today}

\begin{abstract}
We theoretically investigate the anisotropic dispersion features of a two dimensional electron gas at the (110) oriented $\mathrm{LaAlO}_3/\mathrm{SrTiO}_3$ interfaces, as revealed by scanning gate microscopy of electronic flow from a quantum point contact. The dispersion relation of the (110) $\mathrm{LaAlO}_3/\mathrm{SrTiO}_3$ interface is characterized by a highly non-circular Fermi surface. Here, we develop an efficient tight-binding model for the electron gas at the interface. We show that the anisotropy of the Fermi surface causes both the direction of the electron flux from the quantum point contact and the periodicity of the self-interference conductance fringes to depend strongly on the orientation of the constriction relative to the crystal lattice. We show that the radially non-uniform distribution of the Fermi velocity on the Fermi surface results in skewing of electron trajectories when the quantum point contact gates are not aligned with the in-plane primitive vectors. We show that this effect results in the separation of electrons belonging to different orbitals for wide (110) $\mathrm{LaAlO}_3/\mathrm{SrTiO}_3$ quantum wells.
\end{abstract}

\maketitle

\section{Introduction}
Two-dimensional electron gases (2DEGs) created at the interfaces between band insulating transition metals such as $\mathrm{LaAlO}_3$ (LAO) and oxides such as $\mathrm{SrTiO}_3$ (STO) \cite{Ohtomo2004} have recently gained considerable attention due to the rich phenomena they exhibit, ranging from magnetism to superconductivity \cite{Hwang2012}. The possibility of electrostatic control of the potential landscape at the 2DEG level in oxide interfaces, comparable to that in conventional 2DEGs embedded in semiconducting heterostructures, has facilitated the experimental realization of a few-electron quantum dots \cite{PhysRevMaterials.4.122001, Prawiroatmodjo2017}, promising for all-electrical spin manipulation \cite{PhysRevB.109.155306, PhysRevApplied.22.044012}. Electrostatic gating was also exploited for the implementation of a split-gate quantum point contact (QPC) \cite{Thierschmann2018}, which enabled the observation of conductance quantization, previously obtained in lithographically etched \cite{doi:10.1126/science.1168294, PhysRevB.94.045120} LAO/STO waveguides under substantial external magnetic fields \cite{Annadi2018}, evidencing ballistic electron transport \cite{Jouan2020}. 

The most commonly studied 2DEG in oxide heterostructures is the one created at the LAO/STO interface grown in [001] direction. The band structure of this interface, in the low-energy regime, is dominated by a parabolic and isotropic titanium $d_{xy}$ band separated by $50 - 300$ meV from highly anisotropic $d_{yz/xz}$ bands \cite{Pai_2018, Gariglio_2019}. Recently, a high mobility 2DEG has been realized at the LAO/STO (110) interface. Interestingly, it was found that the electron transport properties are highly anisotropic when comparing the two in-plane crystallographic directions \cite{Ariando_NatCom}. A similar effect was demonstrated for the case of SrTiO$_3$ (110) without a LaAlO$_3$ layer but covered by a two-dimensional tetrahedrally coordinated titania overlayer \cite{Diebold_PNAS}. In that case, ARPES measurements show two elliptical Fermi surfaces rotated with respect to each other by 90 degrees. It was also proposed that electron diffraction can be used to elucidate the anisotropic dispersion in two-dimensional electron systems \cite{PhysRevB.110.085424}.  

Here, we demonstrate an alternative method for studying the anisotropy of the Fermi surface based on the Scanning Gate Microscopy (SGM) of electron flow from a split-gate QPC in an (110) LAO/STO interface. Scanning gate microscopy is a technique that enables spatial imaging of electron flow in semiconductor nanostructures. A negatively charged atomic force microscope scans above the surface of the structure, and due to capacitive coupling to the 2DEG, a small depletion region beneath the tip is created, which deflects the electron trajectories. Simultaneous measurement of the conductance allows the detection of regions with high electron flow. The seminal SGM mapping of the electron flow from QPC revealed that electrons escape the constriction in a fan-shaped flux, and that the conductance is imprinted with interference fringes \cite{doi:10.1126/science.289.5488.2323, Topinka2001, PhysRevLett.105.166802, 10.1063/1.1484548}. The fringes result from the self-interference of electrons escaping the QPC constriction and being scattered back by the SGM tip \cite{PhysRevB.80.041303, PhysRevLett.94.126801}. The fringe separation depends on the Fermi wavelength, which can be controlled by a back-gate \cite{10.1063/1.1484548}. The SGM technique has been widely used in recent years for the visualization of magnetic focusing of electron trajectories \cite{Aidala2007, Bhandari2016, PhysRevB.98.115309}, edge and valley physics in monolayer materials 
\cite{Mrenca_2015, Prokop_2020}
or electron-hole states in normal-superconducting hybrids \cite{Bhandari2020, PhysRevB.106.035432, PhysRevB.109.115410, maji2025scanninggatemicroscopydetection}. 

In this paper, we demonstrate that the electron flow probed by the SGM technique reveals unique features characteristic of the anisotropic dispersion of the 2DEG formed at the (110) oriented LAO/STO interface. For this purpose we develop an efficient tight-binding model for the LAO/STO interface, which we utilize to numerically calculate the SGM conductance maps. We observe that the focusing of the fan-like electron flow, the separation of the interference fringes, and the orientation of the flow itself depend on the orientation of the QPC with respect to the crystal lattice. We show how these features are related to the anisotropic electron Fermi wavelength and Fermi velocity.

The paper is structured as follows. In Sect. II, we introduce the tight-binding model used for the transport calculations. Section III A presents the conductance quantization and SGM conductance maps corresponding to three orientations of the QPC. In Sect. III B, we discuss the case of an impurity deflecting the electron trajectories, and in Sect. III C, we present the results for a wide well in the growth direction, where the three bands are populated. Section IV provides the discussion, and Sect. V summarizes the paper.

\section{Theory}

\subsection{Tight-binding model}
\begin{figure}[ht!]
\center
\includegraphics[width = 0.9\columnwidth]{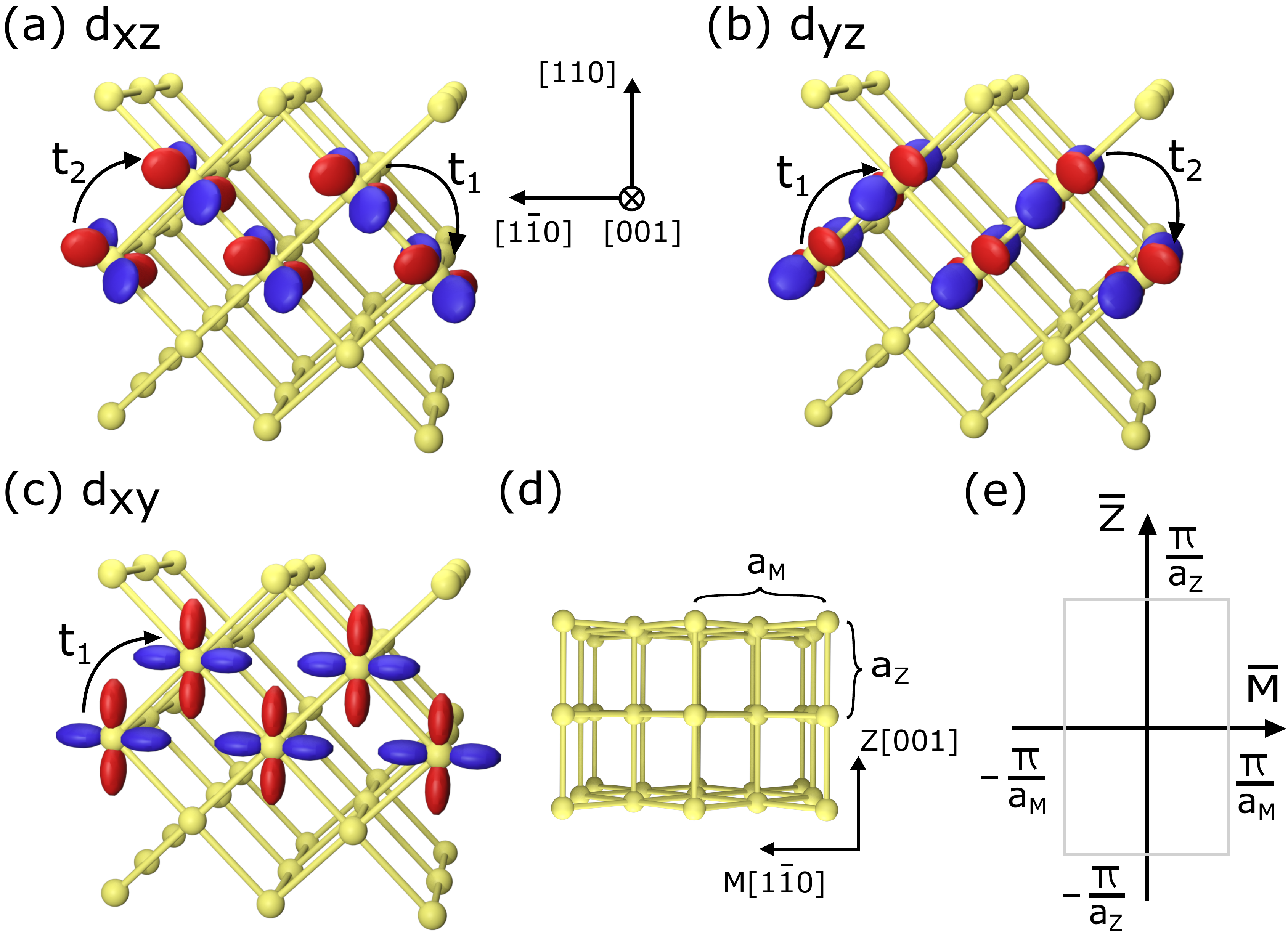}
\caption{Schematic representation of the (110) oriented SrTiO$_3$ structure together with the nearest neighbor hoppings between the $d_{xz}$ (a), $d_{yz}$ (b), and $d_{xy}$ (c) orbitals of the Ti lattice sites. (d) The top view of the stucture with two (110) in-plane lattice constants $a_M$ and $a_Z$. (e) The (110) in-plane Brillouin zone in the momentum  space.}
\label{fig:lattice_110}
\end{figure}

Following Ref. \cite{Diebold_PNAS} we start by considering $N$ TiO$_2$ atomic layers stacked one next to the other along the [110] direction of the SrTiO$_3$ substrate. These layers correspond to the region where the electrons are confined due to the polarity discontinuity at the LAO/STO interface. The minimal model describing such a system can be constructed based on the three $t_{2g}$ orbitals of the Ti sites \cite{Ariando_NatCom,Diebold_PNAS,MacDonald_001_PRB}. For the [110] orientation, the Ti atoms form a rectangular lattice with hoppings $t_1$ and $t_2$ between the nearest neighbors (cf. Fig. \ref{fig:lattice_110}). After diagonalization, the corresponding tight-binding Hamiltonian in momentum space has the following form
\begin{equation}
    \hat{H}=\sum_{kln\sigma} E^{l}_{\mathbf{k}, n} \hat{a}_{\mathbf{k}ln\sigma}^{\dagger}\hat{a}_{\mathbf{k}ln\sigma},
    \label{eq:hamiltonian_0}
\end{equation}
where
\begin{equation}
\begin{split}
    E^{xy}_{\mathbf{k}, n} &= -4t_1 - 2t_2+2t_2\cos{k_Za_z} \\
    &-2 \left[  2t^2_1+2t^2_1\cos{(k_Ma_M)} \right]^{1/2}\cos{\left(\frac{\pi n}{N+1}\right)},\\
    E^{yz/xz}_{\mathbf{k}, n} &= -4t_1 - 2t_2+2t_1\cos{k_Za_Z} \\
    &-2 \left[  t^2_1+t^2_2+2t_1t_2\cos{(k_Ma_M)} \right]^{1/2}\cos{\left(\frac{\pi n}{N+1}\right)},\\
\end{split}
\label{eq:disp_rel_110}
\end{equation}
and $xy$, $yz$, and $xz$ correspond to the three $t_{2g}$ orbitals at each lattice site, $N$ to the number of layers, and $n$ indexes the quantized energy levels that arise from confinement in the $[110]$ direction. The DFT calculations show that the inter-orbital terms are either zero or negligible, and the hopping parameters have the values $t_1=-0.277$ eV, $t_2=-0.031$ eV \cite{Diebold_PNAS} leading to a degenerate $xz$ and $yz$ band. The two components of the momentum vector $\mathbf{k}=(k_Z,k_M)$ correspond to the two directions in momentum space defined by in-plane reciprocal lattice vectors, while $a_M=a$, $a_Z=\sqrt{2}a$, where $a=0.39$ nm is the distance between two Ti nearest neighboring lattice sites (cf. Fig. \ref{fig:lattice_110}). 

\begin{figure}[ht!]
\center
\includegraphics[width = 0.9\columnwidth]{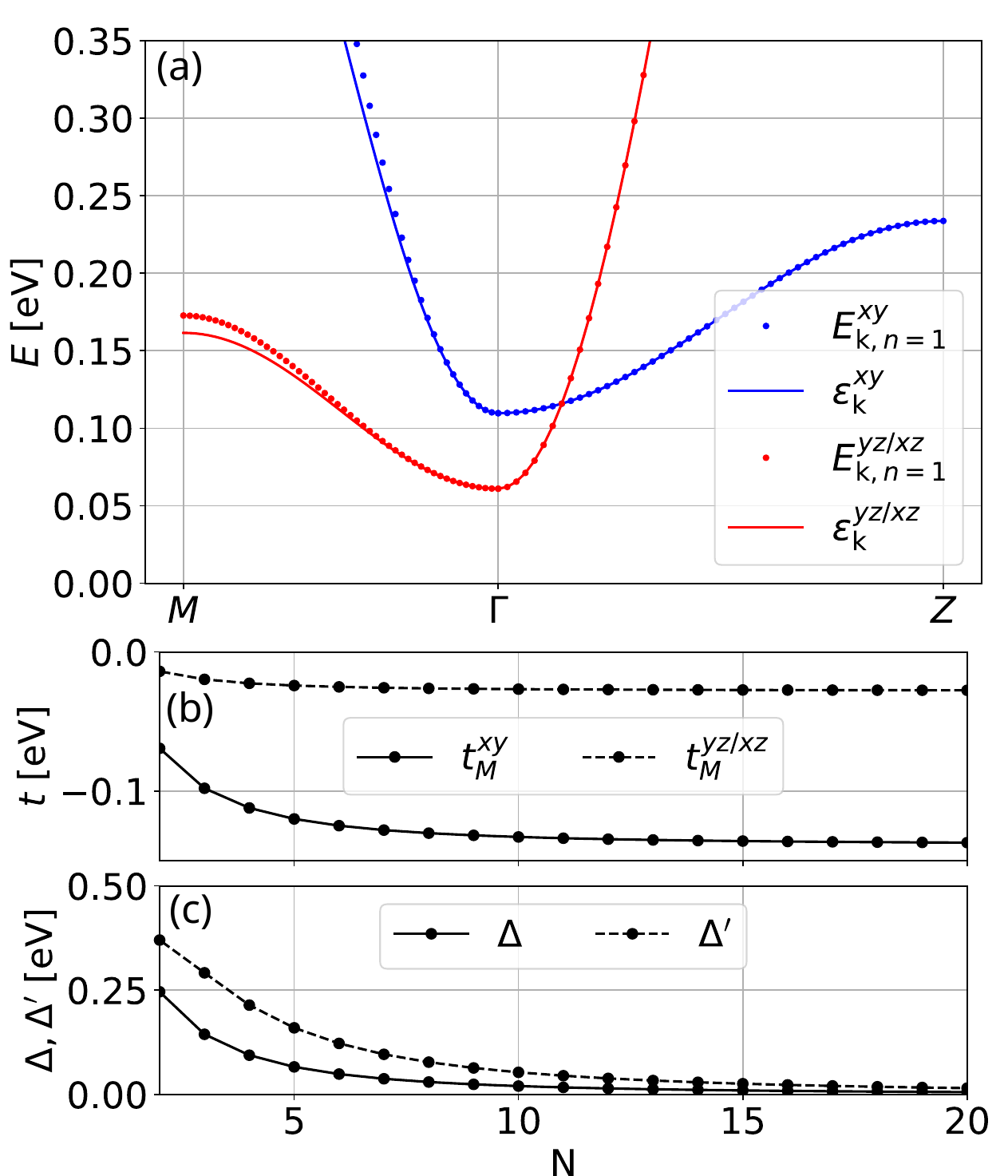}
\caption{(a) Low-energy bands of the (110) LAO/STO interface plotted along the high symmetry directions. The solid lines correspond to the effective lattice model given by Eq. (\ref{eq:hamiltonian_eff}) while the dots correspond to the original dispersion relations determined by Eq. (\ref{eq:disp_rel_110}) for $n=1$ with $N=6$; (b) The effective hopping parameters as a function of the number of Ti layers in the [110] direction; (c) The $N$-dependence of the band offset between the $xy$ and $yz/xz$ bands for $n=1$ [$\Delta$ defined by Eq. (\ref{eq:band_offset})] as well as the energy difference between the first $yz/xz$-excited state and the $xy$-ground state [$\Delta'$ defined by Eq. (\ref{eq:excited_ground_state_energy})] of the [110] quantization.}
\label{fig:bands_fitting}
\end{figure}

To carry out electron transport simulations of a QPC nanodevice with typical dimensions, a relatively large number of lattice sites must be included in our system. Consequently, an approximation must be introduced in order to perform the calculations in a reasonable time. Since the considered system is characterized by a very low electron concentration, we can simplify the situation by focusing on the lowest-lying energy state for which $n=1$. Additionally, the dispersion relations given by Eq. (\ref{eq:disp_rel_110}) for $n=1$ can be well described by an effective lattice model of the following form
\begin{equation}
\begin{split}
    \hat{H}_{\mathrm{eff}}&=\sum_{il\sigma} \big(\;t^l_{M}\;\hat{a}_{i\pm\hat{r}_M,l,\sigma}^{\dagger}\hat{a}_{i,l,\sigma} + t^l_{Z}\;\hat{a}_{i\pm\hat{r}_Z,l,\sigma}^{\dagger}\hat{a}_{i,l,\sigma}\\
    &+\epsilon_{0}^l\;\hat{a}_{i,l,\sigma}^{\dagger}\hat{a}_{i,l,\sigma}\;\big),
\end{split}
\label{eq:hamiltonian_eff}
\end{equation}
where the $t^l_{Z}$ and $t^l_{M}$ are the hopping energies between the nearest neighbor sites of a two-dimensional rectangular lattice with the lattice vectors $\hat{r}_M=(a_M,0)$ and $\hat{r}_Z=(0,a_Z)$. It is convenient to express the onsite energy $\epsilon^l_0$ in a following form
\begin{equation}
    \epsilon^l_0=-2t_M^l-2t_Z^l+\delta_l,
\end{equation}
where $\delta_l$ corresponds to the minima of the resulting bands. Note that the effective model corresponds to a two-dimensional rectangular lattice, and the actual thickness of the 2DEG in the [110] direction is taken into account via the hopping parameters, as we show later. After transforming the effective Hamiltonian to reciprocal space, one obtains
\begin{equation}
 \hat{H}_{\mathrm{eff}}=\sum_{\mathbf{k}l\sigma}\epsilon_{\mathbf{k}l}\hat{a}^{\dagger}_{\mathbf{k}l\sigma}\hat{a}_{\mathbf{k}l\sigma},
 \label{eq:Hamiltonian_start_110}
 \end{equation}
where
\begin{equation}
\begin{split}
    \epsilon_{\mathbf{k}l} &= 2t^l_{Z}\big(\cos{(k_Za_Z)}-1\big)\\
    &+ 2t^l_{M}\big(\cos{(k_Ma_M)}-1\big)+\delta_l.\\
    \end{split}
\label{eq:disp_rel_110_eff}
\end{equation}
In order to map the original dispersion relations described by Eq. (\ref{eq:disp_rel_110}) for $n=1$ onto the effective model, we set $t_Z^{xy}=t_2$ and $t_Z^{yz/xz}=t_1$ and apply a fitting procedure with respect to $t_M^{xy}$, $t_M^{yz/xz}$, and $\delta_l$. For an exemplary value of $N$, the original bands defined by Eq. (\ref{eq:disp_rel_110}) and the effective ones corresponding to Eq. (\ref{eq:disp_rel_110_eff}) are provided in Fig. \ref{fig:bands_fitting}. 
As one can see, the effective model well reconstructs the anisotropic behavior of the original dispersion relations, which results from the symmetry of the $t_{2g}$ orbitals with respect to the confinement direction. Another principal feature of the band structure is the appearance of the band offset between the $xy$ and $yz/xz$ bands, which can be expressed as
\begin{equation}
    \Delta = E^{xy}_{\mathbf{k}=(0,0),n=1}-E^{yz/xz}_{\mathbf{k}=(0,0),n=1}.
    \label{eq:band_offset}
\end{equation}

By applying the back-gate voltage, one can tune the Fermi energy \cite{Shen2016}, which allows accessing both the single- or two-band regime in the considered system as proposed in Ref. \cite{singh2018gap}. It should be noted that both the anisotropic effective hopping parameters and the band offset depend significantly on the thickness of the 2DEG only in the range $N\approx1-10$ as shown in Fig. \ref{fig:bands_fitting} (b) and (c). Moreover, the first $yz/xz$ excited state of the [110] quantization is always above the $xy$-ground state, meaning that $\Delta'>0$, where 
\begin{equation}
    \Delta' = E^{yz/xz}_{\mathbf{k}=(0,0),n=2}-E^{xy}_{\mathbf{k}=(0,0),n=1}.
    \label{eq:excited_ground_state_energy}
\end{equation}
Hence, in principle one can always tune the Fermi energy in such a manner that the $xy$ and $yz/xz$ bands originating from the ground state of the quantization in the [110] direction are occupied, leaving the excited states empty. As the thickness of the 2DEG in (110) LAO/STO has been estimated to range between a few nanometers \cite{singh2018gap} to a dozen nanometers \cite{Han_2014} in the Results Section, we will consider two cases, with a narrow and a wide 2DEG.

For a relatively low thickness of 2DEG at the surface of (110) oriented STO, both ARPES measurements and DFT/tight-binding calculations reported in Ref. \cite{Diebold_PNAS} show a band structure similar to that considered here in Fig. \ref{fig:bands_fitting}. Namely, in both cases, the bottom of the yz/xz band is shifted with respect to the xy band by about $30-50$~meV, and the values of the effective masses resulting from our model and those obtained theoretically and experimentally in Ref. \cite{Diebold_PNAS} are in relatively good agreement, as shown in Table \ref{tab:effective mass}.

\begin{table}
    \centering
    \begin{tabular}{ccccc}
        \hline
       & $m_Z^{xy}$ & $m_Z^{yz/xz}$ & $m_M^{xy}$ & $m_M^{yz/xz}$\\
       \hline
       Theor. $m^*$ & $8.1$ & $0.9$ & $1.0$ & $5.0$\\
       Theor. $m^*$ from Ref.\cite{Diebold_PNAS} & $8.2$ & $0.6$ & $0.6$ & $4.7$\\
         Expt. $m^*$ from Ref.\cite{Diebold_PNAS}& $9.7$ & $0.67$ & $0.74$ & $6.1$\\
         \hline
    \end{tabular}
    \caption{Effective mass in the units of electron mass along the two high symmetry directions ($M$ and $Z$) for the $xy$ and $yz/xz$ bands as calculated with the use of the considered tight-binding effective model with $N=6$ (first row). For comparison we provide both the theoretical and experimental values from Ref. [15] (second and third row).}
    \label{tab:effective mass}
\end{table}

It should be noted that the effective model given by Eq. (\ref{eq:hamiltonian_eff}) is defined on a rectangular lattice with lattice constants: $a_M=\sqrt{2}a$, $a_Z=a$, where $a=0.39$ nm. This imposes a further limitation when it comes to simulations of devices of realistic size. To overcome this limitation, we apply a scaling procedure to the effective Hamiltonian, which allows us to reduce the number of lattice sites without reducing the size of the system \cite{wojcik2024scaledtightbindingmodel}. Namely, we rescale our lattice constants by taking: $\tilde{a}^{\prime}_M=sa_M$, $\tilde{a}_Z=sa_Z$, where $s$ is the scaling parameter. As a consequence, the hopping parameters also have to be modified accordingly: $\tilde{t}^l_Z=t_Z^l\;(a_Z/\tilde{a}_Z)^2$, $\tilde{t}^l_M=t_Z^l\;(a_M/\tilde{a}_M)^2$. 

As a final step, we supplement our model with the quantum point contact potential and the potential arising from the presence of the SGM tip. The resulting Hamiltonian takes the form
\begin{equation}
    \begin{split}
    \hat{H}^{(s)}_{\mathrm{eff}}&=\sum_{il\sigma}\big(\;\tilde{t}^l_{x}\;\hat{a}_{i\pm\hat{x},l,\sigma}^{\dagger}\hat{a}_{i,l,\sigma} + \tilde{t}^l_{y}\;\hat{a}_{i\pm\hat{y},l,\sigma}^{\dagger}\hat{a}_{i,\sigma}\big)\\
    &+\sum_{il\sigma}(\tilde{\epsilon}_0^l-\varepsilon_0+V_i)\;\hat{n}_{il\sigma},
    \end{split}
\label{eq:final_tb_model}
\end{equation}
where, for the sake of clarity, we have replaced the notation corresponding to the two original directions $M$ and $Z$ within the (110) plane with $x$ and $y$, respectively. Additionally, we have included the energy shift $\varepsilon_0$ which allows us to set the bottom of the lowest band at zero energy for all values of $N$, while $V_i$ is defined in the following manner
\begin{equation}
    V_i=V_{QPC}(x_i,y_i)+V_{SGM}(x_i,y_i),
\end{equation}
where $V_\mathrm{QPC}(x,y)$ models the split-gate QPC potential \cite{Davies} 
\begin{align}
\frac{V_{\mathrm{QPC}}(x,y)}{V_g}=&\frac{1}{\pi}[\arctan(\frac{w+x'}{d})+ \arctan(\frac{w-x'}{d})]\nonumber\\
&-g(s+y',w+x')-g(s+y',w-x')\nonumber\\
&-g(s-y',w+x')-g(s-y',w-x'),
\end{align}
where 
\begin{equation}
    g(u,v)=\frac{1}{2\pi}\arctan(\frac{uv}{dR}),
\end{equation}
and $R=\sqrt{u^2+v^2+d^2}$. The primed coordinates are related to the real-space coordinates as
\begin{align}
    & x' = (x - x_s)\cos(\theta) - (y - y_s)\sin(\theta)\nonumber\\
    & y' = (x - x_s)\sin(\theta) + (y - y_s)\cos(\theta),
\end{align}
which enables us to position the QPC with the center of the constriction at $(x_s, y_s)$ and rotated by the angle $\theta$. 

The potential $V_{SGM}(x,y)$ induced by the SGM tip at the level of the 2DEG is typically of Gaussian or Lorentzian form \cite{PhysRevB.62.5174, PhysRevB.77.125310, PhysRevLett.99.136807, Pala_2009}. We adopt the latter potential distribution following Ref. \cite{PhysRevB.84.075336} 
\begin{equation}
V_{\mathrm{SGM}}(x,y)=\frac{V_{\mathrm{tip}}}{1+\frac{(x-x_{\mathrm{tip}})^2+(y-y_{\mathrm{tip}})^2}{d_{\mathrm{sgm}}^2}}.
\end{equation}

For concreteness, we take the QPC parameters $s = 10$ nm, $w = 30$ nm, and $d = 10$ nm, which are comparable to recent experiments on oxide QPCs \cite{Jouan2020}. For the SGM tip, we select $d_\mathrm{sgm} = 5$ nm and $V_\mathrm{tip} = 0.1$ eV. These parameters yield representative conductance quantization and the SGM variation of conductance comparable to those obtained experimentally. The particular choice of the SGM potential does not affect the phenomena presented herein.

\subsection{Electron transport calculations}
The zero-temperature conductance is obtained through the Landauer-Büttiker formula
\begin{equation}
    G = \frac{2e^2}{h}\sum_{i,j}T_{ij},
\end{equation}
where $T_{ij}$ is the transmission probability for the electrons incoming from the input lead $i$ before the QPC and captured in the lead $j$ after escaping the QPC. The probabilities are derived from the scattering matrix of the system calculated using the Kwant package \cite{Christoph}. The conductance maps are obtained using the Adaptive numerical library \cite{Nijholt2019}. The implementation of the tight-binding model, along with all the code used for the calculations, is available in an online repository \cite{code}.

\begin{figure}[ht!]
\center
\includegraphics[width = 0.8\columnwidth]{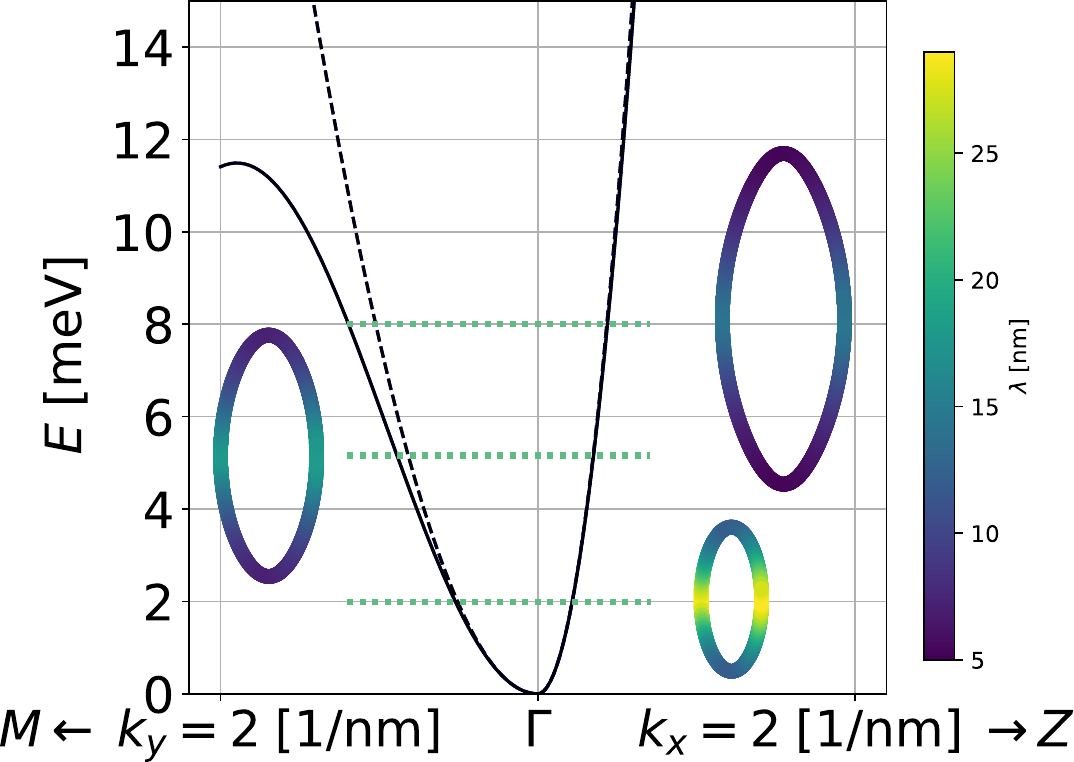}
\caption{The dispersion relation in the vicinity of $\Gamma$ point obtained for the tight binding model Eq. (\ref{eq:final_tb_model}) without scaling procedure ($s=1$, dashed lines) and with scaled model $s = 3$ (solid lines) for a strong confinement in the [110] direction. The plot shows three Fermi surfaces calculated for energies denoted by the corresponding dotted green lines. The colors of the points in the surfaces correspond to the Fermi wavelength.}
\label{fig:LAO_DR}
\end{figure}

\section{Results}

\subsection{Anisotropic band structure}
Let us begin by examining the band structure of the introduced tight-binding model in the limit of strong confinement along the [110] direction. We fit the tight-binding parameters for $N = 6$ (resulting in a thickness of $L = 1.7$ nm), exclude the potentials $V_\mathrm{QPC}$ and $V_\mathrm{SGM}$ from the Hamiltonian Eq. (\ref{eq:final_tb_model}) (implying spatial invariance of the lattice) and set the energy shift $\varepsilon_0 = 0.0609$ eV such that the conduction band minimum is at zero energy.

In Fig.~\ref{fig:LAO_DR}, we present the low-energy band structure of the (110) LAO/STO layer around the $\Gamma$ point. The dashed curve corresponds to the bands obtained without the scaling procedure, i.e. $s = 1$, while the solid curve represents the dispersion calculated with the scaling factor $s = 3$. We observe that at low energies, there is a very good agreement between the two models. The second subband, which also exhibits anisotropic dispersion, is separated by approximately 50 meV from the bottom of the first band. Since we are interested in the low-energy properties, we will work within a Fermi energy regime that allows for the occupation of only the lowest energy subband.

In Fig. \ref{fig:LAO_DR} we depict three Fermi surfaces calculated for $E_F = 2, 5$ and $8$ meV, with colors representing the Fermi wavelength $\lambda = 2\pi/|\mathbf{k}|$. The Fermi contours are distinctly non-circular, exhibiting a significant variation in the Fermi wavelengths along the contour. In the following, we will consider the case of $E_F = 5$ meV. The concrete choice of the Fermi energy has no qualitative effect on the results presented here, and it only affects the Fermi wavelength, so by that the obtained distance between the conductance fringes.

\subsection{SGM conductance maps}
\begin{figure}[ht!]
\center
\includegraphics[width = 0.9\columnwidth]{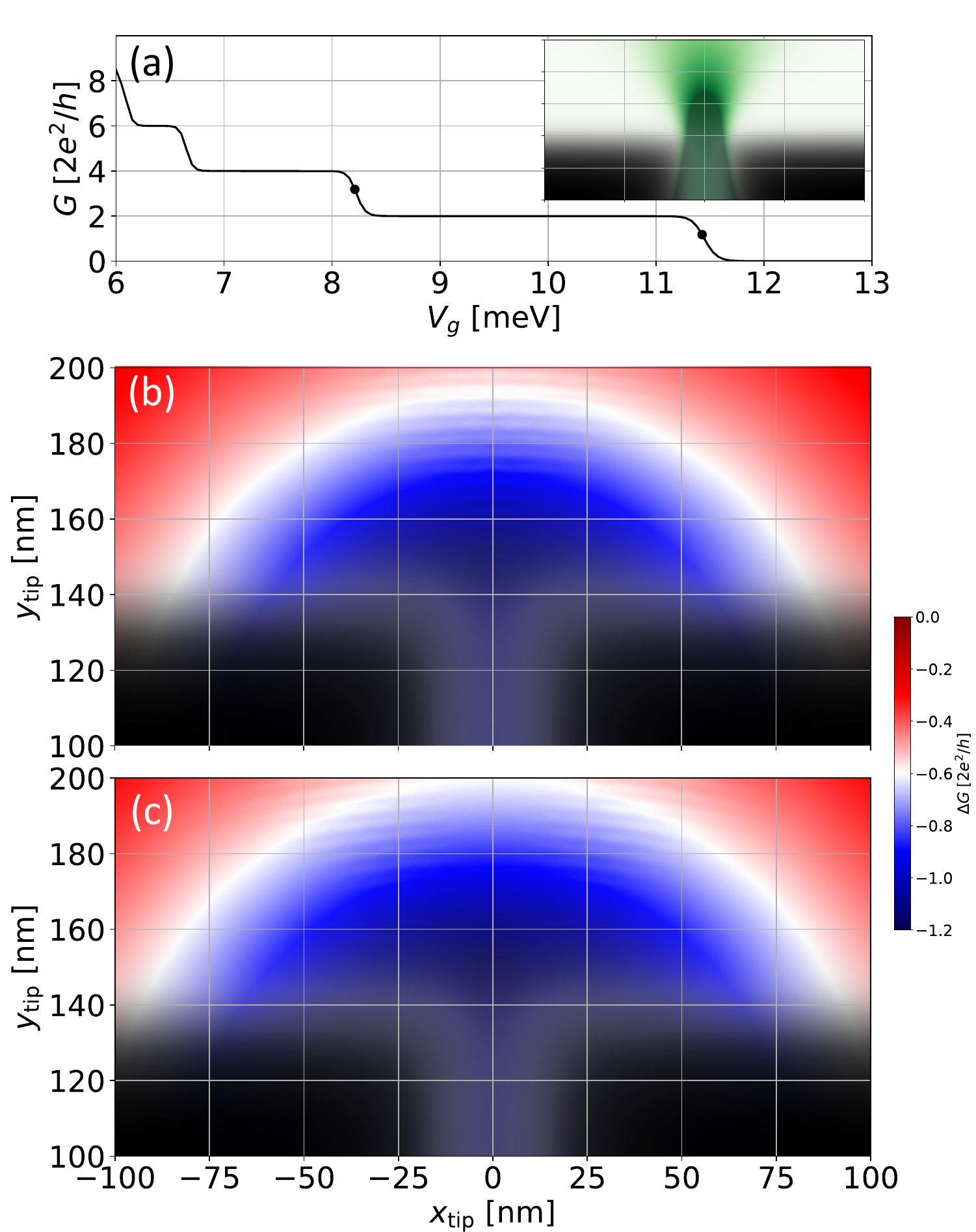}
\caption{(a) Conductance versus the QPC potential $V_g$. The inset shows the probability density obtained for the first conductance step. The range of $x$ and $y$ in the inset is the same as in panels (b) and (c). (b, c) Conductance change map versus the SGM tip potential position for the (b) first and (c) second conductance step [denoted with black dots in (a)]. The QPC is oriented horizontally and its potential is denoted with faint black colors in the maps in (b) and (c).}
\label{fig:LAO_SGM_horizontal}
\end{figure}

We first consider a horizontally oriented QPC with $y_s = 100$ nm and $\theta = \pi/2$. The system consists of a finite scattering region connected to a semi-infinite lead at the bottom (from which the electrons are injected), top, and on the left and right sides of the system above QPC (where the electrons are captured to simulate open boundary conditions). Figures \ref{fig:LAO_SGM_horizontal} (b) and (c) in gray depict the distribution of the QPC potential in the system.

The conductance versus the QPC potential $V_g$ is shown in Fig. \ref{fig:LAO_SGM_horizontal}(a). We observe pronounced conductance plateaus quantized in units of $4e^2/h$ due to spin and $d_{yz/xz}$ band degeneracy. We set $V_g$ at the first two conductance steps (see black dots in Fig.\ref{fig:LAO_SGM_horizontal}(a)) and calculate the conductance of the system versus the SGM tip position $(x_\mathrm{tip}, y_\mathrm{tip})$. In Figs. \ref{fig:LAO_SGM_horizontal} (b) and (c) we plot the difference between the conductance with the tip and that calculated without the tip present ($\Delta G$). In the conductance maps, we observe a strong reduction in conductance as the tip moves closer to the constriction. This is due to a steep slope of the conductance steps, which indicates that a small modification of the potential near the QPC can significantly alter the transmission through the constriction. Importantly, on the map, we also observe faint single- and two-lobe patterns, which correspond to the tip blocking the flow in the first and second subbands, respectively. The lobes are imprinted by fringe patterns resulting from the self-interference of electrons, with the fringe separation $l = \lambda/2$ estimated to be 3.6 nm.

\begin{figure}[ht!]
\center
\includegraphics[width = 0.9\columnwidth]{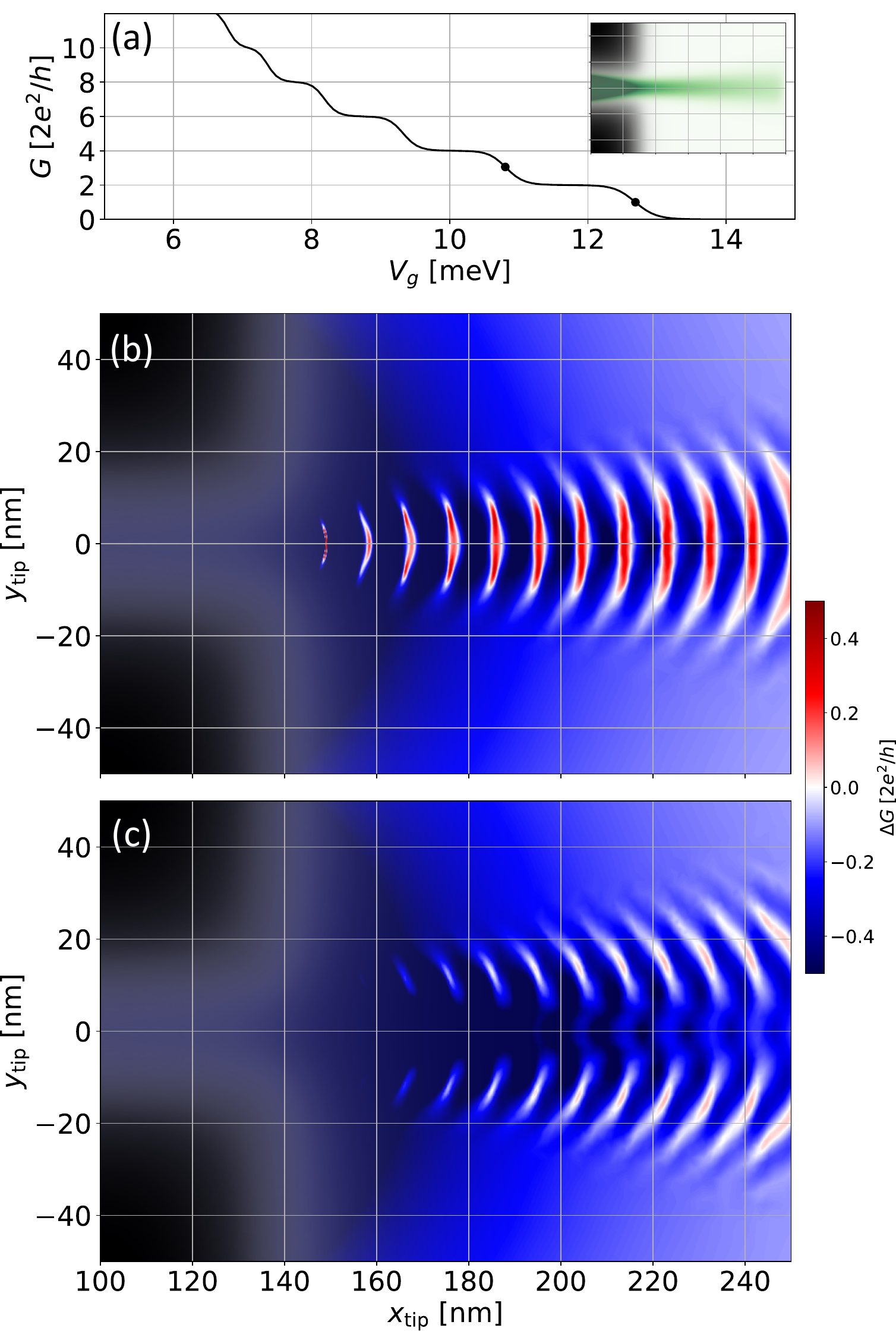}
\caption{Same as Fig. \ref{fig:LAO_SGM_horizontal} but for vertical orientation of the QPC gates.}
\label{fig:LAO_SGM_vertical}
\end{figure}

Now we consider a case with the QPC potential oriented vertically, with $x_s = 100$ nm and $\theta = 0$. Similarly to the horizontal orientation, we place semi-infinite leads before and after the QPC. The conductance trace versus the $V_g$ potential shown in Fig.~\ref{fig:LAO_SGM_vertical}(a) displays more densely packed conductance plateaus with smoother steps compared to the case of Fig.~\ref{fig:LAO_SGM_horizontal}(a). To understand this behavior, let us approximate the low energy (110) LAO/STO dispersion by an effective mass model $H = \hbar k_x^2/2m_x + \hbar k_y^2/2m_y$ with $m_x < m_y$. The energy of the quantized bands in the QPC saddle-point constriction is proportional to $1/m_{x/y}$, with the $x/y$ direction being the direction across the QPC slit. It is evident now that for the case shown in Fig. \ref{fig:LAO_SGM_horizontal}(a), where the effective mass in the direction across the QPC constriction is smaller compared to the case presented in Fig. \ref{fig:LAO_SGM_vertical}(a), the conductance steps are significantly more separated in $V_g$, which results from a considerable separation of the bands in the constriction. On the other hand, in Fig. \ref{fig:LAO_SGM_vertical}(a) the steps are closer to each other due to a large $m_y$, but also exhibit a more gentle slope as the QPC constriction becomes shorter compared to the Fermi wavelength $\lambda \propto 1/\sqrt{m_{x/y}}$.

The $\Delta G$ maps shown in Figs. \ref{fig:LAO_SGM_vertical}(b) and (c) depict the single- and two-lobe SGM patterns, along with the conductance fringes. They are separated by $l = \lambda/2 = 9.1$ nm, a considerably larger distance compared to that observed in Figs. \ref{fig:LAO_SGM_horizontal}(b) and (c), due to the significantly larger Fermi wavelength in the $x$ direction (see the inset to Fig. \ref{fig:LAO_DR}).

In Figs. \ref{fig:LAO_SGM_vertical}(b) and (c) we observe that the conductance oscillations exhibit a significantly greater amplitude for a vertically oriented QPC. In this case, the flow from the QPC is more focused---compare the insets in Figs. \ref{fig:LAO_SGM_vertical}(a) and \ref{fig:LAO_SGM_horizontal}(a). For vertical orientation, the spread at a distance of 100 nm from the center of the QPC is less than 40 nm, while for horizontal orientation it is approximately 100 nm. Consequently, the electron flow is more influenced by the tip, as a significantly larger portion of the total wavefunction is scattered by the SGM tip. Another remarkable feature of the conductance maps is the non-circular shape of the conductance fringes, i.e., they are more flat (convex) in the horizontal orientation of the QPC Fig. \ref{fig:LAO_SGM_horizontal}(a,b) (vertical orientation of the QPC Fig. \ref{fig:LAO_SGM_vertical}(a,b)). This results from the dependence of the Fermi wavelength on the direction of electron propagation, which translates into angular change of the separation between the conductance fringes.

\begin{figure}[ht!]
\center
\includegraphics[width = 0.9\columnwidth]{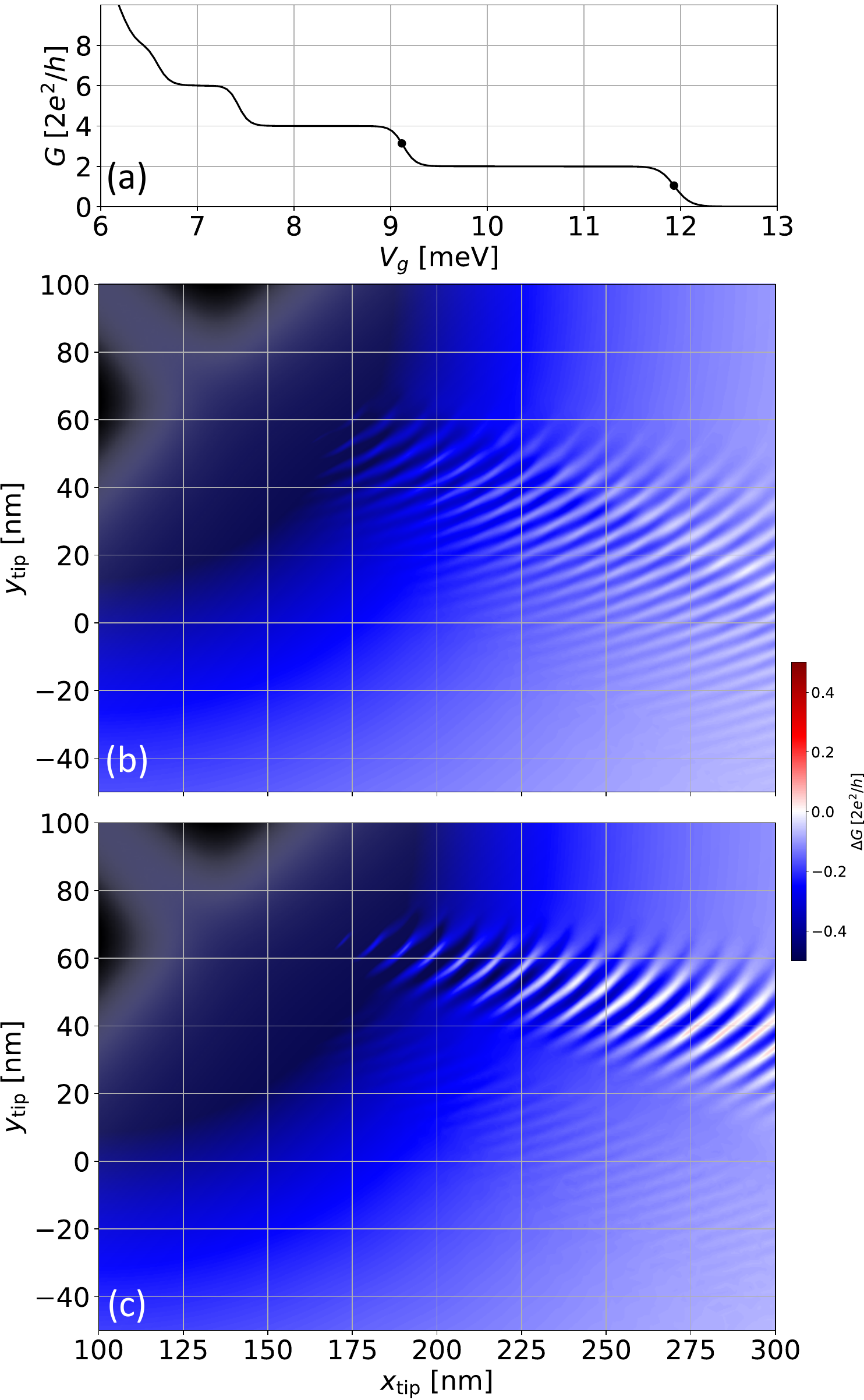}
\caption{(a) Conductance versus the QPC $V_g$ potential. (b) and (c) conductance change map versus the SGM tip potential position for the (a) first and (b) second conductance step [denoted with black dots in (a)]. The QPC is oriented diagonally and its potential is denoted with faint black colors in the maps in (b) and (c).}
\label{fig:LAO_SGM_diagonal}
\end{figure}

\begin{figure}[ht!]
\center
\includegraphics[width = 0.8\columnwidth]{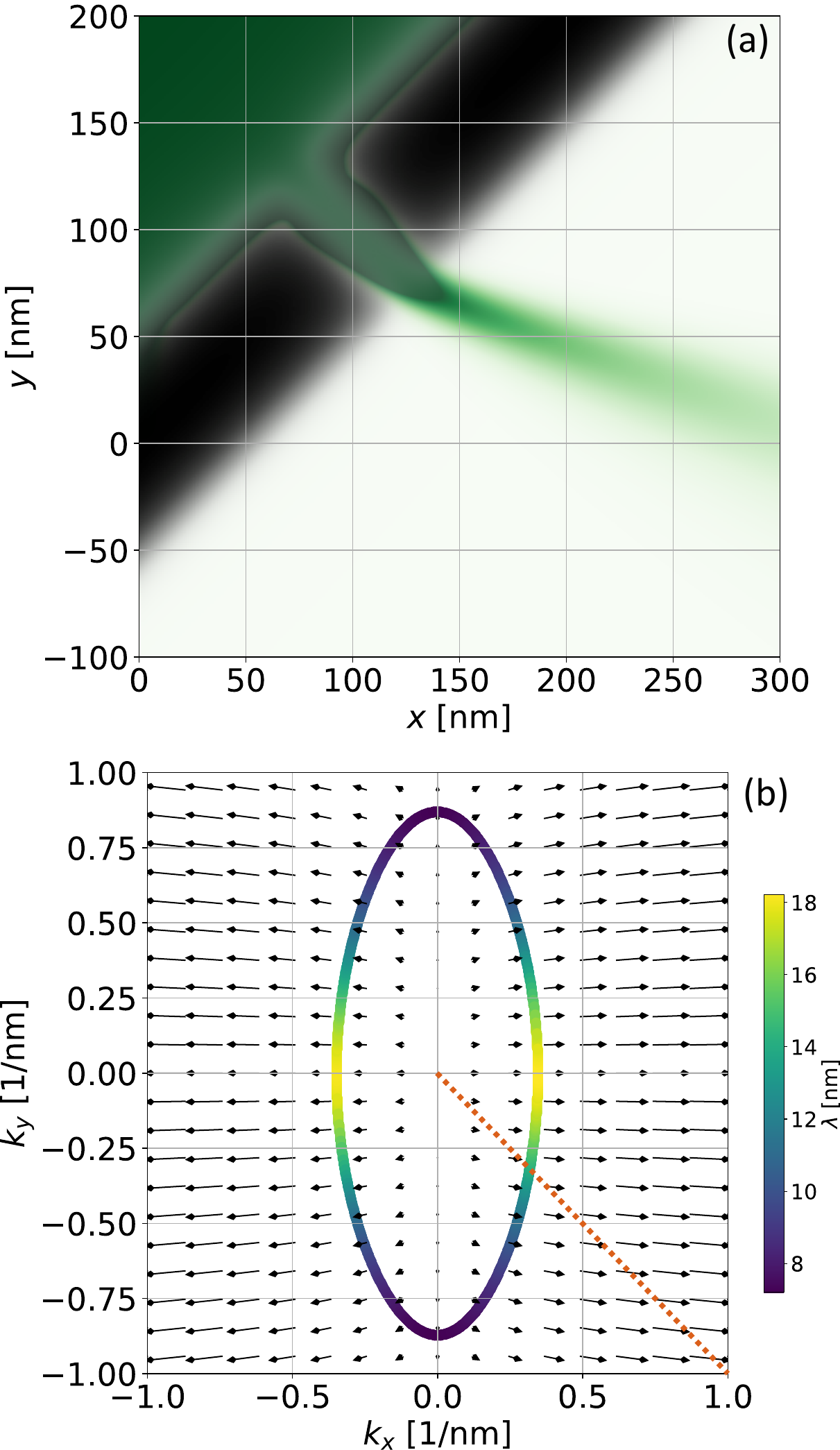}
\caption{(a) Density probability (green) of an electron flow from a QPC (with the potential denoted with gray) at the first conductance step for diagonal orientation of the QPC gates. (b) The arrows show the group velocity obtained from the band structure. The colors show the Fermi wavelength at the Fermi contour for $E_F=0$.}
\label{fig:LAO_velocities}
\end{figure}

The most interesting case arises when the QPC electrodes are not aligned with any of the in-plane primitive vector directions. We consider a scenario in which the QPC electrodes are diagonally oriented with $\theta = \pi/4$, and the center of the constriction is located at $x_s = 100$ nm and $y_s = 100$ nm. In Fig. \ref{fig:LAO_SGM_diagonal}(a) we present the conductance as a function of the $V_g$ potential, and in Figs. \ref{fig:LAO_SGM_diagonal}(b) and (c) the corresponding conductance maps obtained for the QPC potential tuned to the first and second conductance steps, respectively. A remarkable change in the separation of the fringes is visible in the $\Delta G$ maps. For the features along the horizontal line, the fringe separation is large, while for the electrons propagating further down the system, the separation significantly decreases. This variation represents the distribution of the Fermi wavelength across the Fermi surface—see the inset of Fig. \ref{fig:LAO_DR} and Fig. \ref{fig:LAO_velocities}(b).

Another surprising feature is the visible shift of the single and double lobes shown in Figs. \ref{fig:LAO_SGM_diagonal}(b) and (c) from the diagonal line extending from the QPC slit. This can be better observed in Fig. \ref{fig:LAO_velocities}(a) obtained without the SGM tip potential. There, in gray, we denote the QPC gate potential, and in green, the probability density distribution for the QPC tuned to the first conductance step. The electrons escaping the constriction do not follow the $x=-y$ line pointing from the constriction; rather, they are bent upwards, which translates into the skewed lobes observed in Figs. \ref{fig:LAO_SGM_diagonal}(b) and (c). This phenomenon can be understood by inspecting the Fermi velocities. In Fig. \ref{fig:LAO_velocities}(b) we show the Fermi contour along with the Fermi velocities calculated numerically from the energies given by the tight-binding model Eq. (\ref{eq:final_tb_model}) with $\mathbf{v} = [\partial E/\partial k_x, \partial E/\partial k_y ]/\hbar$. We clearly observe that the $x$ component of the velocities is dominant as $m_x < m_y$, as also evidenced by the effective mass model approximating the velocities as $\mathbf{v} = \hbar[k_x/m_x, k_y/m_y]$. In the QPC constriction, the electron wavefunction takes the form of a standing wave between the QPC electrodes and the expectation value of the momentum in the $x = y$ direction is $\langle k_x + k_y \rangle = 0$. Upon escaping the constriction and entering the open region, the point $k_x + k_y = 0$ on the Fermi surface corresponds to a higher velocity in the $x$ direction compared to the $y$ direction [see the orange dashed line in Fig. \ref{fig:LAO_velocities}(b)]. As a result, the particle follows a trajectory oriented towards the positive $y$ direction when $m_x < m_y$. 

\subsection{Impurity scattering}
In the experimentally studied devices, the electron stream is influenced by impurities present in the system, which can separate the original current fans into electron branches propagating in different directions \cite{Topinka2001, Jura2007}. In each branch, the electrons propagate coherently, and the branches are decorated by interference-induced fringes. In the system with symmetric dispersion, the fringes maintain consistent spacing \cite{PhysRevB.94.075301}. In the case of (110) LAO/STO, impurity scattering opens up the possibility for a further study of anisotropic electron dispersion. 

\begin{figure}[ht!]
\center
\includegraphics[width = 0.9\columnwidth]{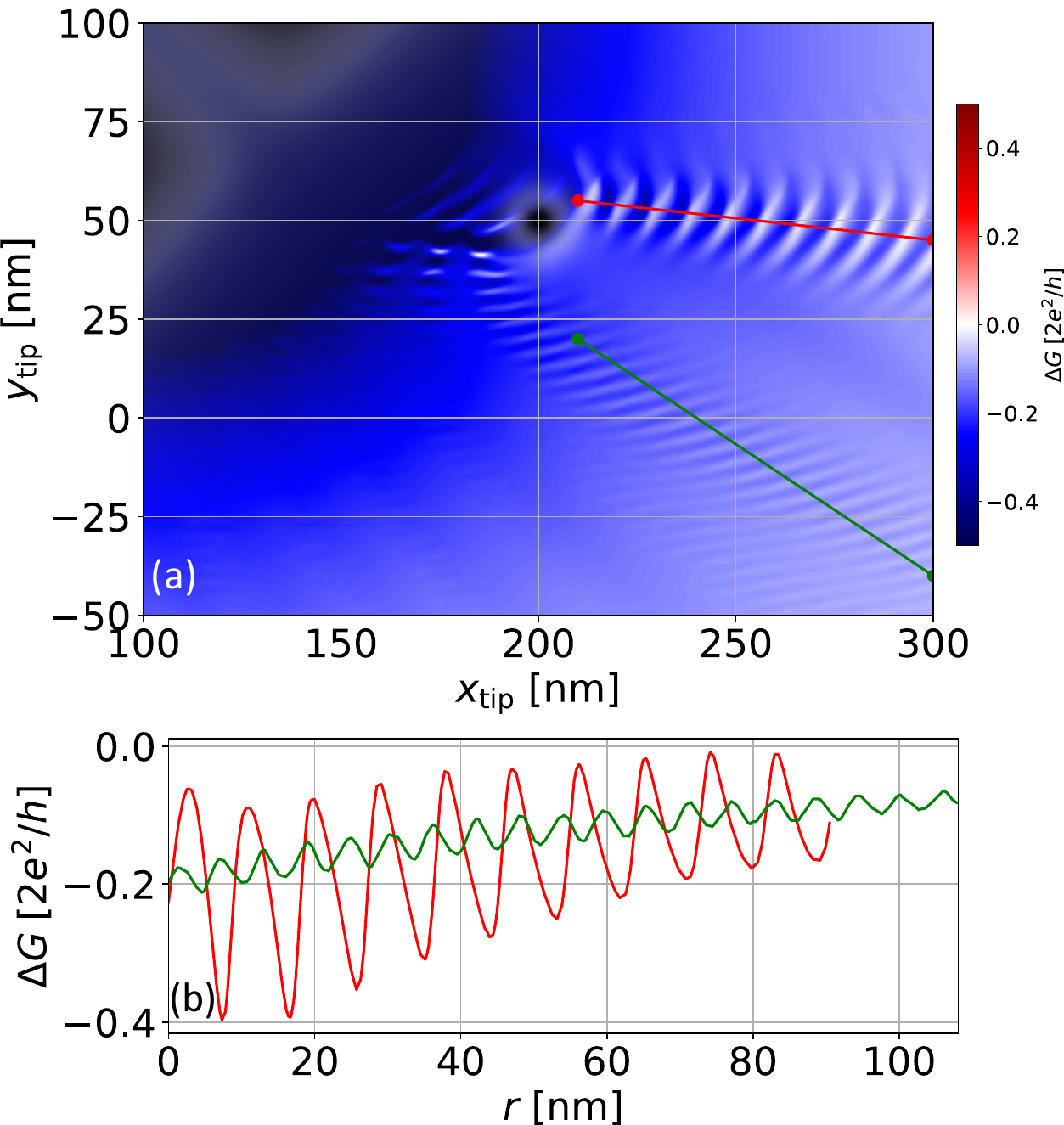}
\caption{(a) Conductance change map for diagonal orientation of the QPC with a scattering Gaussian potential placed in front of the QPC opening. The gray colors denote the potential of the QPC and the Gaussian impurity. (b) Conductance cross-sections along two lines denoted in the map (a)}
\label{fig:LAO_gaussian}
\end{figure}

We now consider an example case of a single impurity. In Fig. \ref{fig:LAO_gaussian}(a) we present a SGM conductance map obtained for the diagonally oriented QPC tuned to the first conductance step. We extend the potential term $V_{i}$ in the tight-binding model Eq. \ref{eq:final_tb_model} by introducing a Gaussian potential $V_s(x,y) = A\exp(-(x-x_0)^2/2\sigma^2) - (y-y_0)^2/2\sigma^2)$ with $A = 5$ meV, $\sigma = 5$ nm and $x_0 = 200$ nm, $y_0 = 50$ nm. The QPC potential and the scattering center are visualized by shades of gray in Fig. \ref{fig:LAO_gaussian}(a).

We observe that the scattering center splits the otherwise single fan of electron flow into two streams with distinctly different propagation directions. The difference in the direction of propagation is translated into different Fermi wavelengths, which is reflected in the change of the self-interference fringes visible in the map of Fig. \ref{fig:LAO_gaussian}(a). A more detailed analysis of the conductance oscillations through cross-sections of the conductance maps along the red and green lines is shown in Fig. \ref{fig:LAO_gaussian}(b). In the plot, we observe a higher amplitude of the oscillations for the stream propagating more along the $x$ direction compared to the one with the diagonal stream, which is consistent with the features in the SGM maps for horizontal and diagonal orientations of the QPC. Most importantly, the increase in wavelength for electrons propagating in the $x$ direction is translated into an increase in the conductance oscillation period seen in Fig. \ref{fig:LAO_gaussian}(b) for the red curve compared to the blue one.

\subsection{Weak constriction in [110] direction}
\begin{figure}[ht!]
\center
\includegraphics[width = 0.8\columnwidth]{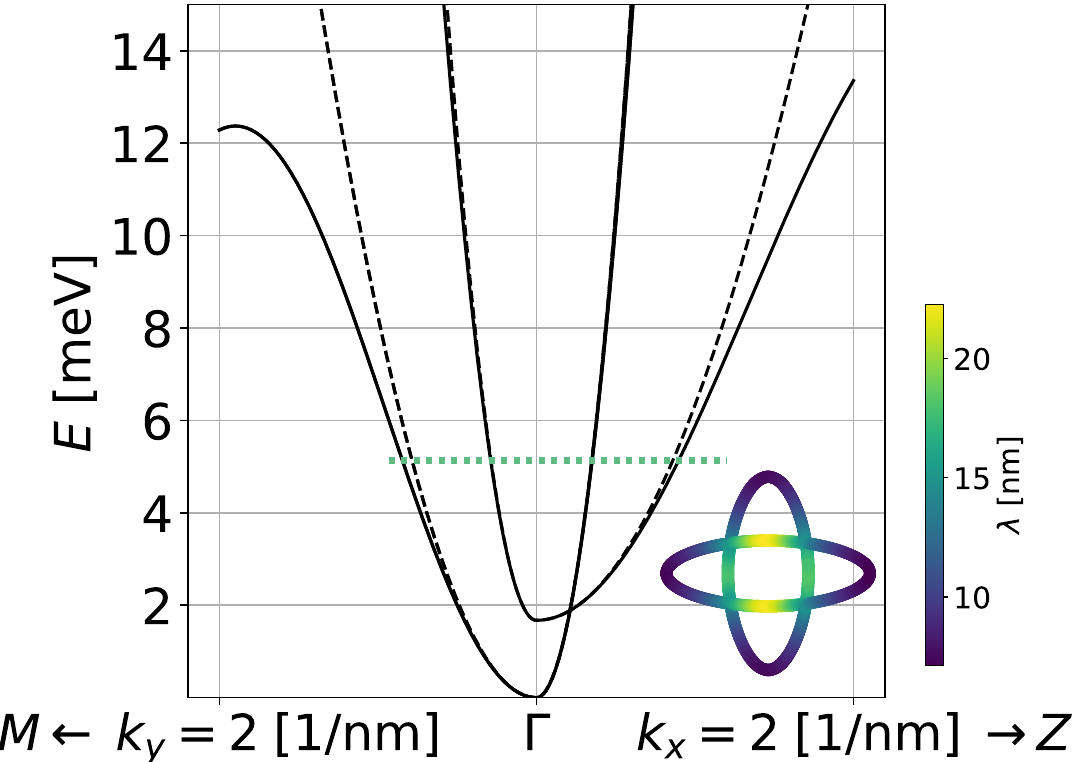}
\caption{The dispersion relation in the vicinity of $\Gamma$ point obtained for the tight binding model Eq. \ref{eq:final_tb_model} without scaling procedure ($s=1$, dashed lines) and scaled model $s = 3$ (solid lines) for a wide confinement in the [110] direction. The plot shows the Fermi surface calculated for $E = 5$ meV. The color points in the surfaces correspond to the Fermi wavelength.}
\label{fig:LAO_DR10nm}
\end{figure}

We now consider a wider quantum well in the [110] direction with $N = 37$, which results in a thickness of $L=  10.2$ nm. We imply spatial invariance of the lattice and set the energy shift $\varepsilon_0 = -0.0021$ eV so that the band bottom is at zero energy as considered previously. For this thickness, the offset between the $xy$ and $yz/xz$ bands is significantly reduced [see the solid curve in Fig. \ref{fig:bands_fitting}(c)]. The band structure is displayed in Fig. \ref{fig:LAO_DR10nm} and the inset shows the Fermi surface at $E_F = 5$ meV, with colors corresponding to the Fermi wavelength. As one can see for $E_F = 5$ meV, both bands contribute to transport, and hence we anticipate observing fringe patterns corresponding to both. However, note that the two bands have opposite orientations of the Fermi surface and we previously saw that the features in the conductance maps related to the fan-shaped electron flow, along with the interference fringes, are mostly visible for the electrons with larger Fermi velocity. This would cause the conductance maps to be dominated mainly by the conductance feature of one of the bands for vertical and diagonal orientations for the QPC---the one with the larger velocity in the direction of the QPC constriction. 

\begin{figure}[ht!]
\center
\includegraphics[width = 0.9\columnwidth]{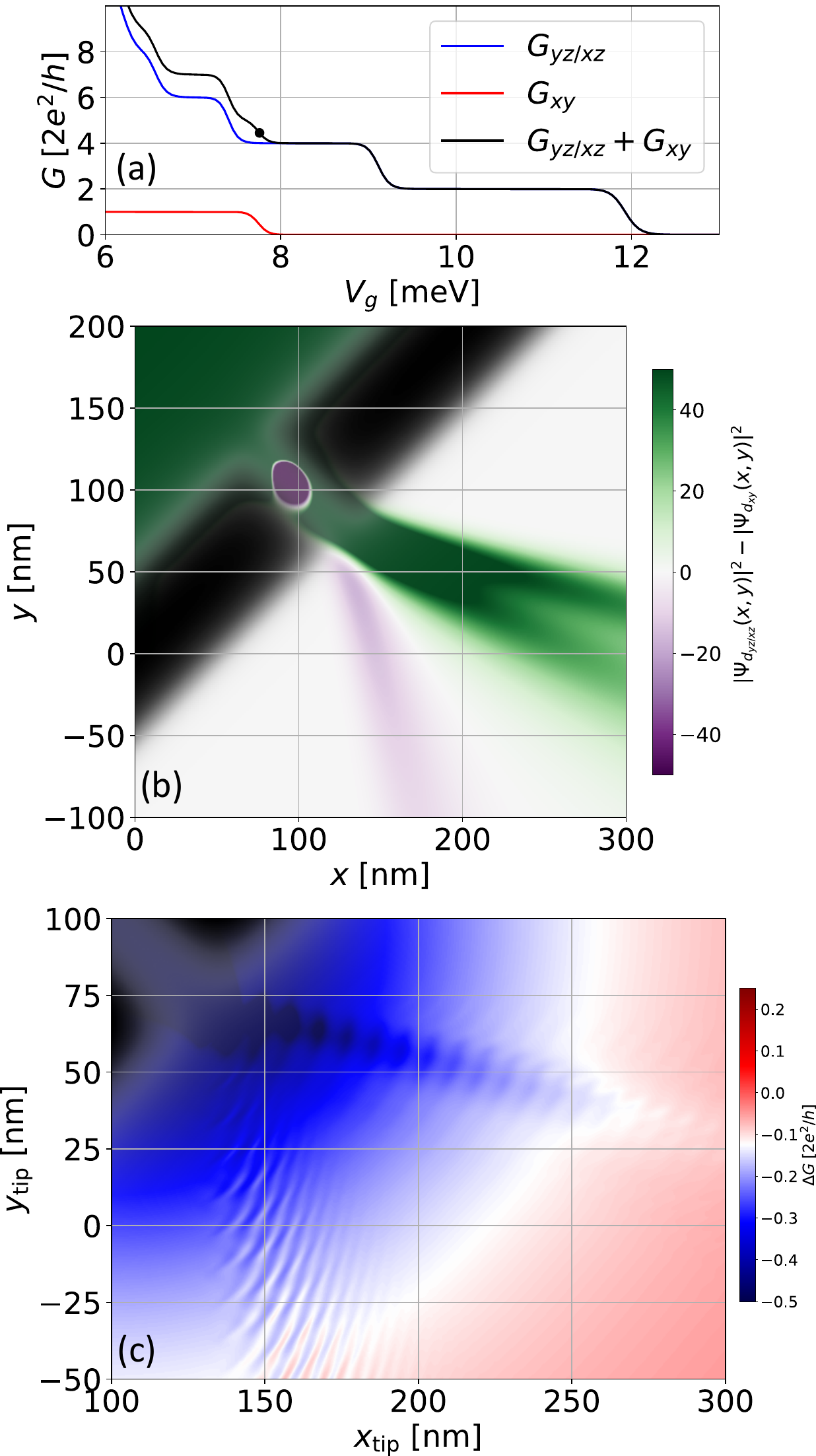}
\caption{(a) Conductance versus the QPC $V_g$ potential. Blue and red colors denote contribution to the conductance from $d_{yz/xz}$ and $d_{xy}$ orbitals respectively. Black curve shows the total conductance. (b) The difference between the probability distribution coming from $d_{yz/xz}$ orbitals and $d_{xy}$ orbital. (c) Conductance change map versus the SGM tip potential position. (b) and (c) are obtained for the $V_g$ value denoted with the black dot in (a).}
\label{fig:LAO_SGM_diagonal_10nm}
\end{figure}

The situation is different for the diagonal orientation of the QPC shown in Fig. \ref{fig:LAO_SGM_diagonal_10nm}. First of all, since the electrons in both bands possess a similar wave vector in the $x=-y$ direction, we anticipate that both bands should contribute to the overall conductance with a comparable length of the conductance steps. In Fig. \ref{fig:LAO_SGM_diagonal_10nm}(a), the blue curve represents the contribution to the conductance from the $d_{yz/xz}$ bands, while the red curve indicates the conductance contribution from the $d_{xy}$ band. The band separation is reflected in the shift of the red conductance trace towards smaller $V_g$ potentials. The black curve depicts the overall conductance. 

Figure \ref{fig:LAO_SGM_diagonal_10nm}(b) shows the difference between the probability densities corresponding to the electron in $d_{yz/xz}$ and $d_{xy}$ bands obtained for the $V_g$ value set to the first conductance step of $d_{xy}$ electrons, at which point the QPC fully transmits the two modes of the $d_{yz/xz}$ band. [see the black dot in Fig. \ref{fig:LAO_SGM_diagonal_10nm}(a)]. Analogously to Fig. \ref{fig:LAO_velocities}(a) we observe a skewed electron flow toward positive $y$ values corresponding to the $d_{yz/xz}$ band, but now with a split at the end of the presented flow due to the occupation of the second subband of the QPC. As the Fermi surface of the $d_{xy}$ band is effectively rotated by $90^\circ$, so is its velocity distribution---the dominant velocity for this band is oriented in the $y$ direction. This results in the skewing of the trajectories of the $d_{xy}$ electrons in a direction opposite to the flow of $d_{yz/xz}$ electrons. This effectively leads to the splitting of the overall electron flow into two main branches, with a spatial separation of the $d_{yz/xz}$ and $d_{xy}$ electrons. The splitting presented in the density map is clearly visible in the SGM conductance map in Fig.~\ref{fig:LAO_SGM_diagonal_10nm}(c) along with the interference fringes.

\section{Discussion}
While in this paper we mainly focus on the (110) LAO/STO interface, in principle, the concept of deducing the anisotropy of the Fermi surface using the SGM technique should be applicable to other systems with non-symmetric Fermi surfaces. LAO/STO interfaces with different orientations, such as (001) also possess anisotropic dispersion \cite{PhysRevB.91.241302}, but only for the $d_{yz/xz}$ bands, which are typically separated by (50-300) meV from the lowest isotropic $d_{xy}$ energy band. In the experiments, only a few first conductance steps belonging to the lowest energy band were observed \cite{PhysRevB.103.235120}. Therefore, to observe the anisotropic effects of the upper bands, the subband quantization in the QPC should be on the order of the band separation, necessitating the use of a few nanometer-wide constrictions to observe flow from the higher band at the first conductance steps. Furthermore, the complicated band structure near the band bottom of $d_{yz/xz}$ orbitals, consisting of avoided crossings, could further hinder the measurement of conductance fringes. Another example of an oxide-based 2DEG that possesses an anisotropic electronic structure has been reported recently for AlO$_x$/KTaO$_3$ \cite{Flavio_2023_AEM}. Beyond the oxides, the anisotropic dispersion occurs also in single and multilayer black phosphorus \cite{PhysRevB.92.075437, Sa_2015, PhysRevB.90.085402, PhysRevB.101.235313, Yuan2015} and their twisted counterparts \cite{PhysRevB.111.075434} however, no demonstration of gated nanostructures has yet appeared in these systems.

It should be noted that the presented calculations were conducted for zero temperature. In the experiments, the distance over which the interference effects can be observed is limited by the thermal excitations of the charge carriers. In the discussed case, the limiting distance is determined by twice the distance between the QPC and the SGM tip positions which allows for self interference. At non-zero temperature, electrons distributed according to the Fermi distribution participate in the transport and become out of phase after traveling over a thermal length \cite{PhysRevB.80.041303} $L_{x/y} = h^2/2\pi m_{x/y} \lambda_{x/y}k_B T$, which for the parabolic band approximation becomes $L_{x/y} = E_F \lambda_{x/y}/\pi k_B T$. For the considered system, at $E_F = 5$ meV for the lowest band with a minimum at zero energy, we obtain $L_{x}/2 = 486$ nm and $L_{y}/2 = 207$ nm, which would allow the observation of dozens of interference fringes [cf. Fig. \ref{fig:LAO_gaussian}] at liquid $^3$He temperatures $T = 350$ mK, at which the SGM measurements can be conducted \cite{PhysRevB.80.041303}.

\section{Summary and conclusions}
We theoretically studied the SGM probing of the electron flow from the QPC realized in (110) LAO/STO interfaces. We developed an effective three-band model that allows for efficient transport calculations for systems with dimensions in the hundreds of nanometers. The anisotropic dispersion in (110) LAO/STO translates to direction-dependent electron velocities and wavelengths. We demonstrated that this results in a significant change in conductance quantization and self-interference effects in a QPC system probed by the SGM technique. Comparing the cases of the QPC gates oriented along two in-plane primitive lattice vectors, we observe a pronounced change in the spread of the electron flow and, most importantly, the self-interference fringe separation, which directly reflects the anisotropy of the Fermi wavelength. As the Fermi velocity does not form a uniform radial field, we observe bending of the electron trajectories for diagonally oriented QPCs, which align with the direction of the Fermi velocity. The SGM conductance maps for the diagonal orientation of the QPC facilitate the visualization of changes in the Fermi wavelength with varying directions of propagation, which is most pronounced for systems with scattering centers. Finally, we demonstrate that for (110) LAO/STO interfaces elongated in the growth direction, where three bands are present in the lower part of the energy spectrum, we observe conductance steps not only from low-lying $d_{yz/xz}$ bands but also from the upper $d_{xy}$ band. The electrons belonging to the two bands have opposite Fermi velocity distributions, causing the separation of the electrons escaping from the QPC into two---band polarized---streams. We showed that the SGM technique allows for uncovering these streams and visualization the two wavelengths for the two bands.

\section{Acknowledgments}
This work is financed by the Horizon Europe EIC Pathfinder under the grant IQARO number 101115190 titled "Spin-orbitronic quantum bits in reconfigurable 2D-oxides" and partly supported by the program „Excellence
initiative – research university” for the AGH University. We gratefully acknowledge the Polish high-performance computing infrastructure PLGrid (HPC Center: ACK Cyfronet AGH) for providing computer facilities and support within computational grant no. PLG/2025/018486.

\bibliography{references}
\end{document}